\documentclass[aps,prl,twocolumn,floats]{revtex4}
\usepackage{epsfig}
\usepackage{bm}
\usepackage{latexsym}

\begin{document}

\newcommand{\hide}[1]{}
\newcommand{\tbox}[1]{\mbox{\tiny #1}}
\newcommand{\half}{\mbox{\small $\frac{1}{2}$}}
\newcommand{\sinc}{\mbox{sinc}}
\newcommand{\const}{\mbox{const}}
\newcommand{\trc}{\mbox{trace}}
\newcommand{\intt}{\int\!\!\!\!\int }
\newcommand{\ointt}{\int\!\!\!\!\int\!\!\!\!\!\circ\ }
\newcommand{\eexp}{\mbox{e}^}
\newcommand{\bra}{\left\langle}
\newcommand{\ket}{\right\rangle}
\newcommand{\EPS} {\mbox{\LARGE $\epsilon$}}
\newcommand{\ar}{\mathsf r}
\newcommand{\im}{\mbox{Im}}
\newcommand{\re}{\mbox{Re}}
\newcommand{\bmsf}[1]{\bm{\mathsf{#1}}} 


\title{Quantum Reversibility: Is there an Echo?}

\author{
Moritz Hiller$^{1}$, 
Tsampikos Kottos$^{1}$, 
Doron Cohen$^{2}$ 
and Theo Geisel$^{1}$
}

\affiliation{
$^{1}$Max-Planck-Institut f\"ur Str\"omungsforschung und Fakult\"at Physik
der Universit\"at G\"ottingen, Bunsenstra\ss e 10, D-37073 G\"ottingen, Germany \\
$^{2}$Department of Physics, Ben-Gurion University, 
Beer-Sheva 84105, Israel
}


\begin{abstract}
We study the possibility to undo the quantum mechanical evolution in a time reversal 
experiment. The naive expectation, as reflected in the common terminology (``Loschmidt 
echo"), is that maximum compensation results if the reversed dynamics extends to the 
same time as the forward evolution. We challenge this belief, and demonstrate that the 
time $t_r$ for maximum return probability is in general shorter. We find that $t_r$ 
depends on $\lambda = \varepsilon_{\rm evol}/\varepsilon_{\rm prep}$, 
being the ratio of the error in setting the parameters (fields) for the time reversed 
evolution to the perturbation which is involved in the preparation process.
Our results should be observable in spin-echo experiments where the dynamical 
irreversibility of quantum phases is measured.
\end{abstract}

\maketitle


In this Letter we study the probability of return $P(t_1,t_2)$ for a generalized wavepacket 
dynamics scenario. The system is prepared in some initial state $\Psi_{\tbox{prep}}$, which 
can be regarded as the outcome of a preparation procedure which is governed by a Hamiltonian 
${\cal H}_{\tbox{prep}}$. We assume that the quantum mechanical evolution is generated by
Hamiltonians with classically chaotic limit: The state is propagated for a time $t_1$ 
using a Hamiltonian ${\cal H}_1$, and then the evolution is time-reversed for a time $t_2$  
using a perturbed Hamiltonian ${\cal H}_2$. The corresponding evolution operators are $U_1$ 
and $U_2$. The probability of return to the initial state is 
\begin{eqnarray}
\label{fkernel1}
P(t_1,t_2)= 
|\langle \Psi_{\tbox{prep}} |U_2(t_2)^{-1} U_1(t_1) | \Psi_{\tbox{prep}} \rangle|^2
\end{eqnarray}
There are two special cases that have been extensively studied in the literature. The 
traditional wavepacket dynamics scenario \cite{H91} is obtained if we set $t_2=0$. In 
this context the ``survival probability" is defined as 
\begin{eqnarray}
\label{fkernel2}
P_{\tbox{SR}}(t) \ = \ P(t,0) 
\end{eqnarray}
The ``Loschmidt echo" (LE) scenario is obtained if we set $t_1=t_2=t$. In this context 
the ``fidelity" is defined as 
\begin{eqnarray}
\label{fkernel3}
P_{\tbox{LE}}(t) \ = \ P(t,t) 
\end{eqnarray}
The theory of the fidelity was the subject of intensive studies during the last 3~years 
\cite{P84,JP01,JSB01,CT02,PS02,BC02,WC02,PLU95,VH03}. It has been adopted as a standard 
measure for quantum reversibility following \cite{P84} and its study was further motivated 
by the realization that it is related to the analysis of dephasing in mesoscopic systems 
\cite{Z91}.

In the present Letter we consider the full scenario 
of a time reversal experiment. The probability to find 
the system in its original state is $P(t)=P(t,0)$ 
before the time reversal ($t<T/2$), and 
\begin{eqnarray}
\label{fkernel4}
P(t) \ = \ P(T/2,t-T/2) 
\end{eqnarray}
after the time reversal ($t>T/2$).
The period $T$ is the total time of the experiment.
The naive expectation, which is also reflected in the term ``Loschmidt echo", 
is to have a maximum for $P(t)$ at the time $t=T$. 
We are going to show that this expectation is wrong. 
We find that the maximum return probability 
is obtained at a time $t_r$ which
in general is shorter than that. Namely,
\begin{equation}
\label{tr}
T/2 \le t_r \le T. 
\end{equation}
If we have $t_r=T/2$ we say that there is no reversibility. 
If we have $t_r \sim T$ we say that we have a nearly perfect echo. 
We show that $t_r/T$ is a function of a dimensionless 
parameter $0< \lambda < \infty$. Namely,   
\begin{eqnarray}
\label{sf}
{t_r\over T} = f(\lambda)\quad ;\quad \lambda = \varepsilon_{\tbox{evol}}/\varepsilon_{\tbox{prep}}
\end{eqnarray}
where $\varepsilon_{\tbox{prep}}$ quantifies the difference 
between the evolution Hamiltonian ${\cal H}$ 
and the preparation Hamiltonian ${\cal H}_{\tbox{prep}}$, 
while $\varepsilon_{\tbox{evol}}$ quantifies the difference 
between the two instances ${\cal H}_1$ and ${\cal H}_2$ 
of the evolution Hamiltonian, which are used for the forward 
and for the time-reversed evolution respectively. 
The idea is that there is no way to have a complete control over 
the parameters (fields) of the systems. Therefore there is 
an unavoidable difference ($\varepsilon_{\tbox{evol}}$) 
between these two instances of ${\cal H}$, which by the setup of 
the experiment are regarded as identical. 
The scaling function (\ref{sf}) takes the limiting value $f(0)=1$ (echo) 
while for $\lambda > \lambda^*$ we get $f(\lambda) = 0.5$ (no reversibility). 
Here $\lambda^*$ is some system-specific constant of order unity.

The most popular preparation which is considered in the literature, either in the context of 
wavepacket dynamics or fidelity (LE) studies, is a Gaussian wavepacket. Obviously the choice 
of such preparation is motivated mainly by the wishful thinking of theoreticians. However, in 
many applications, one is not so much interested in evolving an initial Gaussian wavepacket. 
This is certainly the situation in quantum information processing \cite{NC00} and in spin-echo 
experiments \cite{PLU95} where one starts with a {\it random} initial state. Formally a 
Gaussian wavepacket can be regarded as the ground state of a phase-space shifted Harmonic 
oscillator. Therefore it is characterized by a very large $\varepsilon_{\tbox{prep}}$, 
leading to $\lambda \ll 1$. In this Letter we do not assume $\lambda \ll 1$, but rather 
consider the general case. 

In order to develop a general theory we need a model in which we have control over both 
$\varepsilon_{\tbox{prep}}$ and $\varepsilon_{\tbox{evol}}$. We consider a quantized system
whose classical analog has positive Lyapunov exponent. Its Hamiltonian ${\cal H} = {\cal 
H}(Q,P;x)$ depends on a parameter (field) variable~$x$  
which is determined by the experimental setup. For example it can be either a gate voltage or 
a magnetic flux. The dynamics takes place within a classically small (but quantum mechanically 
large) energy window. The classical dynamics is assumed to have a well defined finite correlation 
time $\tau_{\rm cl}$. We consider classically small (but possibly quantum mechanically large)  
perturbations ($x \mapsto x+\delta x$). Accordingly, the Hamiltonian can be linearized as follows: 
\begin{eqnarray}
\label{WBM}
{\cal H} = {\cal E} + \delta x {\cal B}
\end{eqnarray}
We define ${\cal H}_1$ and ${\cal H}_2$ by setting $\delta x =  \pm\varepsilon_{\tbox{evol}}$. 
The requirement of having {\em classically small} $\delta x$ means that 
the phase space structure of ${\cal H}_1$ and ${\cal H}_2$ is similar, 
and that any (small) difference in the chaoticity can be neglected. 
(we have verified that this smallness condition is satisfied for the example below).

The preparation issue requires further discussion. The traditional possibility is to prepare a 
Gaussian wavepacket (also known as a coherent state preparation). We regard this possibility as 
{\em uncontrolled} because the value of  $\varepsilon_{\tbox{prep}}$ is ill defined. To have a 
physically meaningful definition of $\varepsilon_{\tbox{prep}}$ the natural procedure is as 
follows: We define a preparation Hamiltonian ${\cal H}_{\tbox{prep}}$ by setting $\delta x =  
\varepsilon_{\tbox{prep}}$. Then we  start with an initial eigenfunction of ${\cal E}$, and 
evolve it with ${\cal H}_{\tbox{prep}}$ until we get an ergodic-like steady state within an 
energy shell (note \cite{rmrk}). 
The width of this energy shell is proportional to $\varepsilon_{\tbox{prep}}$. 
The resulting wavepacket is used as an initial state for the time reversal experiment. 
For the purpose of comparison we shall consider also a Gaussian wavepacket preparation. 
For such preparation $\lambda \ll 1$ irrespective of its energy width.
The reason is that this preparation does not occupy ergodically its energy shell. 
Formally one can say that for a Gaussian wavepacket ${\cal H}_{\tbox{prep}}$ differs  
enormously from the evolution Hamiltonian. There is no point in quantifying this 
difference. This is the reason why we use our controlled preparation procedure, where 
${\cal H}_{\tbox{prep}}$ differs from the evolution Hamiltonian in a well defined manner. 
In any case we shall verify that a Gaussian wavepacket is indeed 
like taking a preparation with $\lambda \ll 1$.

In our numerical investigation we use the model Hamiltonian
\begin{eqnarray} 
\label{2DW}
{\cal H}(Q,P;x) = \half(P_1^2{+}P_2^2 + Q_1^2{+}Q_2^2) + x Q_1^2 Q_2^2
\end{eqnarray}
with $x=1+\delta x$. It describes the motion of a particle in a 2D well (2DW). The physical 
units are chosen so as to have dimensionless variables. Therefore upon quantization the Planck 
constant $\hbar$ is a dimensionless  quantity. Our numerical study is focused on an energy 
window around $E \approx 3$ where the motion is mainly chaotic with characteristic correlation 
time $\tau_{\rm cl} \approx 1$ \cite{CK01}. The quantization is done with $\hbar=0.012$. We write 
the Hamiltonian matrix as in Eq.~(\ref{WBM}), using a basis such that ${\cal E}$ is diagonal. 
The mean level spacing is $\Delta \approx 4.3 \times \hbar^2$. As expected, on the basis of a 
general ``quantum chaos" argumentation \cite{FP86}, the matrix ${\cal B}$ is a banded matrix. 
More details regarding the band profile can be found in Ref.~\cite{CK01}. The only additional 
piece of information that is needed for the following analysis is the parametric scale $\delta 
x_c$. This is defined as the $\delta x$ which is needed in order to mix neighboring levels. 
It is given by the ratio $\Delta/\sigma$, where $\sigma$  is the root mean square value of 
the near diagonal matrix elements of the ${\cal B}$ matrix. For the above model $\delta x_c 
\approx 3.8*\hbar^{3/2}$.  

Fig.~1 (left panel) displays representative results of simulated time reversal experiments. One 
experiment is done with a coherent state preparation, and we indeed see behavior that looks like 
an echo ($t_r\sim T$). Qualitatively the same behavior is observed for a random preparation that 
has $\lambda \ll 1$. Once we take a preparation with a larger value of $\lambda$, we realize that 
the compensation time is in general $t_r < T$. In particular with the $\lambda > 1$ preparation 
we do not observe any quantum reversibility ($t_r=T/2$). 

\begin{figure}
\epsfig{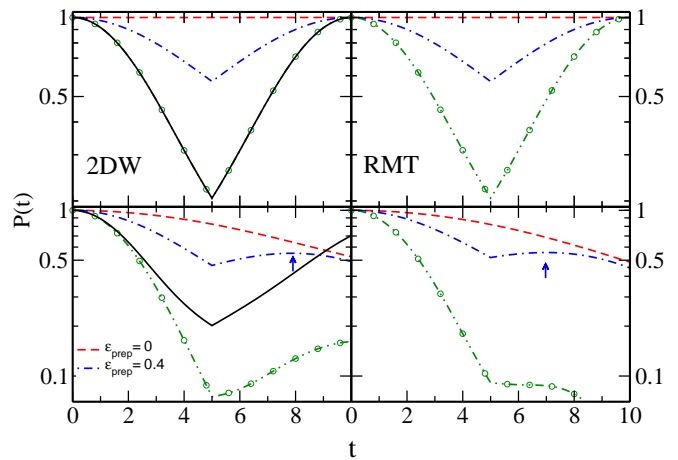}
\caption
{
Left panels:  
The probability $P(t)$ to find the system in its original state, for the 2DW model. 
The two panels, from top to bottom, are for the representative values 
$\varepsilon_{\tbox{evol}}=0.00315$ and $0.26$. 
We use various preparations: a coherent 
state preparation (solid line), a random superposition with the same energy width 
($\circ$), a random superposition (dot-dashed line) with a larger~$\lambda$ where 
$\epsilon_{\tbox{prep}}= 0.4$, and a $\lambda > 1$ preparation (dashed line) which 
was obtained by setting $\epsilon_{\tbox{prep}}= 0$. The arrows indicate the time $t_r$ 
for the case $\epsilon_{\tbox{prep}}= 0.4$. Right panel: The simulations are now done 
with the corresponding RMT model. 
}
\end{figure}

\begin{figure}[b]
\epsfig{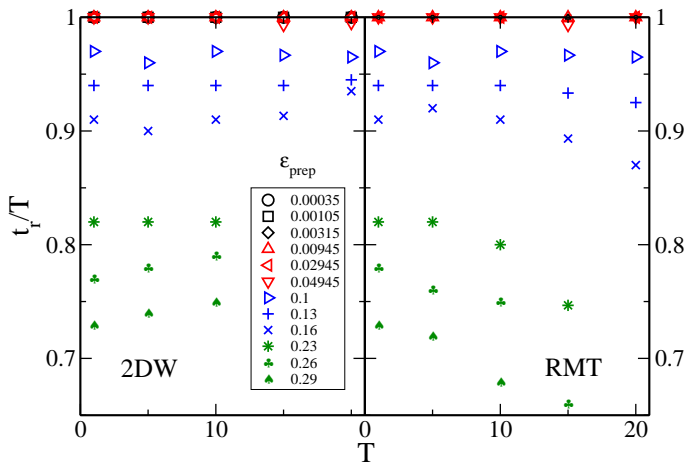}
\caption
{
The time $t_r$ is calculated from the simulation of $P(t)$,  
and the ratio $t_r/T$ is presented for various values of the period $T$, 
and for various values (see legend) of $\varepsilon_{\tbox{evol}}$. 
An average over a number of initial wavepackets 
with the same $\varepsilon_{\tbox{prep}}=0.4$ was performed. We clearly see that $t_r$ 
is smaller for larger values of $\varepsilon_{\tbox{evol}}$. The left panel corresponds 
to the 2DW simulations, while the right panel to the effective RMT model. 
}
\end{figure}

In Fig.~2 we present some results for $t_r/T$. We clearly see that $t_r$ is smaller for 
larger values of $\varepsilon_{\tbox{evol}}$. Much more illuminating is Fig.~3 where we 
present our numerical results for $t_r$ and various $T$. The points corresponding to the 
same $\lambda$ but different values of $\varepsilon_{\tbox{evol}}$ and $\varepsilon_{\tbox{prep}}$ 
fall onto the same smooth curve with a good accuracy, confirming the scaling hypothesis 
(\ref{sf}). Moreover, we see that for $\lambda>\lambda^*$, with $\lambda^* \approx 1.3$, 
we have $t_r=T/2$ (no echo) irrespective of the actual value of $\varepsilon_{\tbox{prep}}$
and $\varepsilon_{\tbox{evol}}$. Thus, if we want a $\lambda > \lambda^*$ type of preparation, 
we simply can set \mbox{$\varepsilon_{\tbox{prep}}=0$}, which means an eigenstate of ${\cal E}$. 
From Fig.~3 we also see that in the other limiting case of $\lambda\ll 1$ we have (nearly) 
an echo. This is the same as in the case of a Gaussian wavepacket preparation. 

\begin{figure}[b]
\epsfig{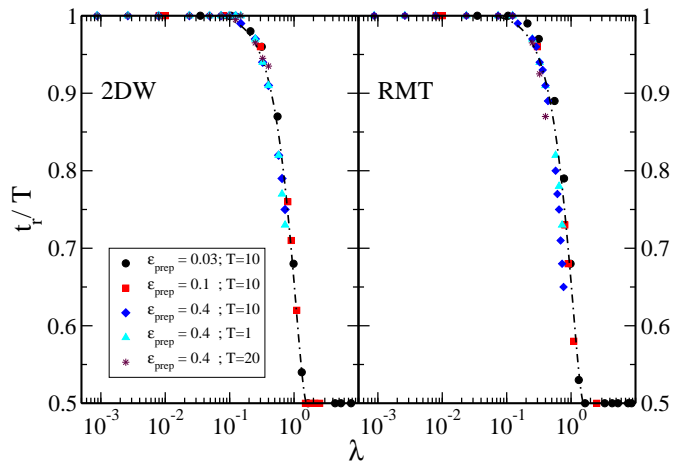}
\caption
{
The ratio $t_r/ T$ for different initial random wavepackets which are determined by 
$\varepsilon_{\tbox{prep}}$,  and for various $\varepsilon_{\tbox{evol}}$. 
The horizontal axis is the scaled variable $\lambda$. 
The data scale nicely in accordance with Eq.~(\ref{sf}). 
The dashed line is a simple polynomial fit. The left panel 
corresponds to the 2DW simulations while the right one to the corresponding RMT model. 
}
\end{figure}

In order to explain the observed results for $t_r$ 
we look on Fig.~4. There are two axes along which 
the theory of $P(t_1,t_2)$ is quite well known. 
One is the traditional wavepacket dynamics axis 
$t_2=0$ along which $P_{\tbox{SR}}(t)$ is defined, 
while the other is the ``LE axis" ($t_2=t_1$) 
along which $P_{\tbox{LE}}(t)$ is defined. 
In the figure we also indicate the course 
of a time reversal experiment.  
It is clear that in order to have $t_r<T$ 
the contor line $P(t_1,t_2)=P_{\tbox{SR}}(T/2)$ 
should meet the $t_1$ axis in a sharp angle. 
Furthermore, if we want to have some remnant of 
an echo at the end of the period ($t=T$)  
we have to cross the contour line  
$P(t_1,t_2)=P_{\tbox{LE}}(T/2)$. 
The condition for that is 
\begin{eqnarray}
P_{\tbox{SR}}(T/2) < P_{\tbox{LE}}(T/2)
\end{eqnarray}
In the following discussion we would like to assume that both $\varepsilon_{\tbox{prep}}$ 
and $\varepsilon_{\tbox{evol}}$ are larger than $\delta x_c$, which means that the 
perturbations are strong enough to mix levels. In such case the decay of $P_{\tbox{SR}}(t)$ 
or $P_{\tbox{LE}}(t)$ is approximately exponential: 
\begin{eqnarray}
\label{expon}
P_{\tbox{SR}}(t), \ P_{\tbox{LE}}(t) \ = \ \exp(-\gamma t)
\end{eqnarray}
Moreover in both cases $\gamma$ is given by the expression 
\begin{eqnarray}
\gamma = \min( \gamma_{\tbox{PT}}, \gamma_{\tbox{SC}} )
\end{eqnarray}
where $\gamma_{\tbox{PT}}$ is the value which is determined by perturbation theory, while  
$\gamma_{\tbox{SC}}$ is determined by semiclassical considerations.   

Once details are concerned the theory behind the exponential approximation Eq.~(\ref{expon}) 
becomes quite different in the two respective cases [$P_{\tbox{SR}}(t)$, $P_{\tbox{LE}}(t)$]. 
The theory of the survival probability is related to the parametric theory of the LDOS 
\cite{CH00,CK01}. Namely, for relatively small perturbations $\gamma = \gamma_{\tbox{PT}} 
\propto (\varepsilon_{\tbox{prep}} / \delta x_c)^2$ is essentially the width of Wigner's Lorentzian, 
while for large perturbations $\gamma = \gamma_{\tbox{SC}} \propto \varepsilon_{\tbox{prep}}$ 
is the width of the energy shell (in the latter case the exponential approximation is at best 
a good fit). In contrast to that, the theory of the fidelity $P_{\tbox{LE}}(t)$ is related to 
a theory of dynamical correlations, and cannot be reduced to the LDOS analysis \cite{WC02}. 
The best theory to date is semiclassical (see ~\cite{VH03} and references therein). As in the 
case of the survival probability we have 
$\gamma = \gamma_{\tbox{PT}} \propto (\varepsilon_{\tbox{evol}}/ \delta x_c)^2$ for small perturbations, while for large perturbations $\gamma = 
\gamma_{\tbox{SC}}$, in some typical cases, is related to the Lyapunov exponent. We would like 
to point out that the study of the general conditions for having a fingerprint of the Lyapunov 
exponent in time reversal experiments is still an open issue for future study \cite{HKC03}. 
The existing semiclassical theory for the ``echo" phenomena assumes Gaussian wavepackets.  
  
By definition $P_{\tbox{SR}}(T/2)$ depends only on $\varepsilon_{\tbox{prep}}$, while 
$P_{\tbox{LE}}(T/2)$ is mainly sensitive to $\varepsilon_{\tbox{evol}}$. Therefore it 
is evident that a small $\lambda$ is a condition for having $t_r<T$. If we have $\lambda
\ll 1$ then the contour line $P(t_1,t_2)=P_{\tbox{LE}}(T/2)$ can be very close to the 
LE-axis. This implies that for $\lambda\ll 1$ we can get nearly 
a perfect echo behavior ($t_r\sim T$).
This picture, as we have seen before, is supported by our numerical findings.
 
\begin{figure}
\epsfig{clip,figure=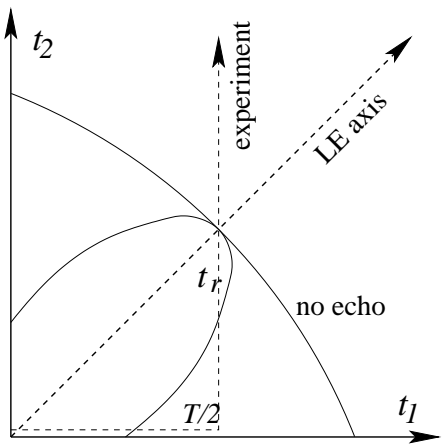,scale=0.6,angle=0}
\caption
{
The contour line of $P(t_1,t_2)$ 
that goes through the point $(T/2,T/2)$. 
One solid line (``no echo") is for the case $\lambda\geq 1$. 
The other solid line is for the case of a relatively 
small~$\lambda$. The dashed line illustrates   
the course of a time reversal experiment, 
while the ``LE axis" is the line along 
which $P_{\tbox{LE}}$ is defined.
}
\end{figure}

As we have seen above, the applicability of semiclassical considerations is not essential 
for having an ``echo". The general picture that we have outlined should be valid also in 
the absence of a semiclassical limit. This is in contrast to the impression that one might 
get from the recent literature. 
In order to establish this provocative statement, in a way that leaves no doubts, 
we use a simple random matrix theory (RMT) procedure. We take the 
resulting banded matrix ${\cal B}$ of the 2DW model~(\ref{2DW}), and randomize the signs 
of the off-diagonal terms. In this way we get an effective RMT (ERMT) model of the type 
that had been introduced by Wigner 50~years ago \cite{wigner}. The model is characterized 
by the same mean level spacing, and by the same band-profile as the physical 2DW model. 
Consequently the generated dynamics is characterized by the same correlation time (the 
latter is determined by the bandwidth). But unlike the 2DW model, the ERMT model is 
lacking a semiclassical limit. In the right panels of figures 1-3 we demonstrate the results 
of simulations that were done with the ERMT model. We clearly see that we get similar 
results (in Fig.3 the RMT drop is slightly sharper).

In summary, we have shed a new light on the physics of quantum reversibility, and in 
particular we have introduced the concept of compensation time~$t_r$, which replaces 
the misleading terminology of ``echo". Our predictions should be tested in wave field 
evolution experiments such as spin polarization echoes in nuclear magnetic resonances
\cite{PLU95,prv}. In particular we have considered the realistic case of a general 
preparation, and clarified the role of semiclassical considerations in the theory. 

\ \\
It is our pleasure to thank Horacio Pastawski and Bilha Segev (BGU) for useful discussions. 
This research was supported by the Israel Science Foundation (grant No.11/02), and by a grant 
from the GIF, the German-Israeli Foundation for Scientific Research and Development.



\begin{thebibliography}{99}
\bibitem{rmrk} To ensure phase space (semiclassical) ergodicity 
of the wavefunction within the energy shell we randomize the signs 
of the state vector elements in the ${\cal H}_0$ basis.
\end{thebibliography}
\end{document}